%Paper: astro-ph/9206004
%From: Phil Fischer <fischer@crocus.Physics.McMaster.CA>
%Date: Mon, 22 Jun 92 22:07:27 EDT

The figures for this paper in postscript form can be obtained through
anonymous FTP as follows.

ftp 130.113.0.111
Name: anonymous
Password: <username>
ftp> cd pub
ftp> get fig1978.ps
ftp> quit

If you have any problems contact fischer@crocus.physics.mcmaster.ca

-----------------------------cut here -----------------

\magnification=\magstep1
\hsize = 6.5 true in\vsize = 9.0 true in
\baselineskip = 20 pt
\overfullrule=0pt
\def\titrule{\vrule height0.4pt depth0.0pt width4.0cm}
\footline={\hfil}
\headline={\ifnum\pageno=1 \hfil \else\hfil\folio \fi}

\newcount\tableone \tableone=0
\newcount\fignum \fignum=0
\newcount\fignumc \fignumc=0
\newcount\equa \equa=0
\newcount\itemm \itemm=0

\font\ftext=cmr10
\font\fsmall=cmr7

\def\taone{\global\advance\tableone by 1 \the\tableone}
\def\taonen{\the\tableone{\ (cont.)}}
\def\itemno{\global\advance\itemno by 1 \the\itemoone}
\def\takeep{\the\tableone}
\def\fig{\global\advance\fignum by 1 \the\fignum}
\def\figcc{\global\advance\fignumc by 1 \the\fignumc}
\def\equat{\global\advance\equa by 1 \the\equa}
\def\itemno{\global\advance\itemm by 1 \the\itemm}

\newdimen\digitwidth\setbox0=\hbox{\rm0}\digitwidth=\wd0
\catcode`~=\active\def~{\kern\digitwidth}
\catcode`!=\active\def!{\kern 0.7\digitwidth}
\catcode`|=\active\def|{\kern 1.4\digitwidth}

\edef\innernewbox{\noexpand\newbox}
\def\sbl{
\innernewbox\tablebox
\centerline { \box\tablebox}
\setbox\tablebox = \vbox \bgroup\offinterlineskip
\halign}

\def\tablerule{\noalign{\hrule}}

\def\sec{${}^{\prime\prime}${ }}
\def\sol{$_\odot$}
\def\et{{\it et al.\ }}
\def\Rc2{{\it R}}
\def\vave{\overline {\hbox{v}}}

\def\tabhl#1#2{{\ftext
\centerline{Table #1}
\centerline{#2}}
\innernewbox\tablebox
\centerline { \box\tablebox}
\setbox\tablebox = \vbox \bgroup\offinterlineskip
\halign\bgroup}

\def\tabhlr#1#2{
\innernewbox\tablebox
%\centerline
{\box\tablebox}
\setbox\tablebox = \vbox \bgroup\offinterlineskip
{\ftext
\centerline{#1}
\smallskip
\centerline{#2}
\medskip
}
\halign\bgroup}

\edef\inskip{\noexpand\newskip}
\def\splr{
\noalign{}
\tablerule\egroup\egroup
\skip3=\wd\tablebox
\skip4=\ht\tablebox
\skip1=0pt
\advance\skip1 by \skip3
\advance\skip1 by -662.2pt
\divide\skip1 by 2
\skip2=0pt
\advance\skip2 by \skip4
\advance\skip2 by 59.5pt
\advance\skip2 by \skip1
\vskip-\skip2
\rotr\tablebox
}

\long\def\tabh#1#2{{\ftext
\centerline{Table #1}
\centerline{#2}}
\innernewbox\tablebox
\setbox\tablebox = \vbox \bgroup
\halign\bgroup}

\long\def\tabhr#1#2{\innernewbox\tablebox
\setbox\tablebox = \vbox \bgroup
{\ftext\hsize=9truein
\centerline{Table #1}
\smallskip
\centerline{#2}
\medskip
\hsize=6.5truein}
\halign\bgroup}

\def\sp{\noalign{\vskip.2truecm}
\tablerule\egroup\egroup
\centerline{\box\tablebox}}

\def\spl{
\noalign{}
\tablerule\egroup\egroup
\centerline {\box\tablebox}}

\def\topper{
\noalign{\vskip0.3cm}
\tablerule
\noalign{\vskip.1cm}
\tablerule
\noalign{\vskip.1cm}
}
\def\topperl{
\noalign{\vskip0.3cm}
\tablerule
\noalign{\vskip.1cm}
\tablerule
}
\def\spacer{
\noalign{\vskip0.2cm}
\tablerule
\noalign{\vskip0.1cm}
}

\newcount\n
\def\hspa#1{}
\def\tcoli{## \hfil}
\def\tcol#1{\hskip#1cm\hfil ## \hfil}
\def\tcolil#1#2{## \hfil\vrule height #1pt depth #2pt}
\def\tcoll#1#2#3{\hskip#1cm\hfil ## \hfil\vrule height #2pt depth #3pt}

\def\vvbs#1#2{\vrule height #1pt depth #2pt}

\pageno=1
\pageinsert
\baselineskip = 15pt
\centerline{\bf DYNAMICS OF THE}
\centerline{\bf INTERMEDIATE-AGE ELLIPTICAL LMC CLUSTER NGC 1978}
\bigskip\bigskip
\centerline{PHILIPPE FISCHER$^{1,2}$, DOUGLAS L. WELCH$^{1,2}$}
\centerline{Department of Physics and Astronomy, McMaster University, Hamilton,
Ontario L8S
4M1, Canada}
\centerline{Email: Fischer@Crocus.Physics.McMaster.CA}
\centerline{ }
\centerline{MARIO~MATEO$^{3}$}
\centerline{The Observatories of the Carnegie Institute of Washington}
\centerline{813 Santa Barbara Street, Pasadena, CA 91101}
\vskip 0.50 truein
\newdimen\addwidth\setbox0=\hbox{Address for proofs: }\addwidth=\wd0
\vskip 0.25 truein
\leftline{Address for proofs:}
\leftline{Philippe Fischer}
\leftline{Department of Physics and Astronomy}
\leftline{McMaster University}
\leftline{1280 Main Street West}
\leftline{Hamilton, Ontario}
\leftline{L8S 4M1  CANADA}
\leftline{Email: Fischer@Crocus.Physics.McMaster.CA}
\vfil
\centerline{To Appear in the Sept. 1992 Astronomical Journal}
\vfil
\line{ \titrule \hfill}
\noindent
$^1$ Guest Investigator, Mount Wilson and Las Campanas Observatories,
which are operated by the Carnegie Institution of Washington.
\centerline{ }
\noindent
$^2$ Visiting Astronomer at the Cerro Tololo Interamerican Observatory,
National Optical Astronomy Observatories, operated by the Association of
Universities for Research in Astronomy Inc., under contract with the
National Science Foundation.
\centerline{ }
\noindent
$^3$ Hubble Fellow.
\endinsert
\vfill\eject

\baselineskip = 20 pt
\centerline{ABSTRACT}
\medskip

BV CCD images of the elliptical LMC cluster NGC 1978 out to a
projected radius R $\sim$ 100 pc were obtained using the 1.0m telescope at
Las Campanas. In addition, radial velocities with a precision of $1.5$
km s$^{-1}$ were measured for 35 member giants using the echelle
spectrographs and 2D-Frutti detectors on the Las Campanas 2.5m and the Cerro
Tololo 4.0m telescopes.

After star-subtraction and median-filtering the ellipticity of the surface
brightness distribution was determined to be $\epsilon = 0.30 \pm 0.03$ and
the major axis position angle to be PA $= 152 \pm 7^\circ$. The stellar
radial velocities indicate that NGC 1978 has a systemic velocity of $\vave =
293.3 \pm 1.0$ km s$^{-1}$.  NGC 1978 appears to be several times older than
its central relaxation time but considerably younger than its half-mass
relaxation time.

Single and multi-mass anisotropic King-Michie models and single-mass
rotating and non-rotating oblate spheroid models were fitted to both the
surface luminosity profiles and the radial velocity data.  The total cluster
luminosity is L$_B = 3.1 - 3.7 \pm 0.2 \times 10^5$ L$_B$\sol\ and L$_V =
3.0 - 3.5 \pm 0.2
\times 10^5$ L$_V$\sol\ (for an assumed LMC distance of 50 kpc), where the
range results from different model extrapolations of the brightness profile.
The multi-mass models, while very effective at constraining the central
mass-to-light ratios (hereafter, M/L's) at about (M/L)$_0$ = 0.13 $\pm$
0.06 M\sol/L\sol, yielded global M/L's which ranged over a factor of 5; M/L
= 0.3 - 1.5 M\sol/L\sol\ for a sample of mass function slopes. The best
agreement between population and dynamical M/L's is seen for the cases $x$ =
0.0 for the B band [(M/L)$_0 = 0.14 \pm$ 0.06 M\sol/L$_B$\sol\ and M/L =
0.35 $\pm$ 0.15 M\sol/L$_B$\sol] and $x$ = 0.5 for the V [(M/L)$_0 =
0.13
\pm$ 0.06 M\sol/L$_V$\sol\ and M/L = 0.40 $\pm$ 0.15 M\sol/L$_V$\sol]. The
single-mass models tended to give better agreement with the luminosity
profiles but produced M/L's (i.e. M/L = 0.20 $\pm 0.08$ M\sol/L$_V$\sol)
that were difficult to reconcile with simple population studies without
invoking a rather high low-mass cut-off (i.e. 0.8 M\sol).

We found no significant differences between the M/L's derived with oblate
spheroid models and those derived with spherical models. While the
non-rotating (anisotropic) models were in better agreement with the
kinematic data, it was impossible to completely rule out the rotating
models. As well, there is no morphological evidence for a merger.

\medskip
\centerline{1. INTRODUCTION}
\medskip

Globular clusters present a unique opportunity to study the internal
dynamics of resolved stellar systems in which the two-body relaxation
timescales are similar to the current ages. One can kinematically examine
clusters to dynamically determine masses and mass-to-light ratios
(hereafter, M/L's) in order to constrain the initial mass function (IMF).
Attempts can also be made to determine the internal cluster dynamics at
various stages in their evolution to try to form a coherent picture of
formation, energy equipartition, and mass segregation.  At later
evolutionary stages, gravothermal catastrophe and the resulting core
collapse can be studied.

In the Milky Way, these studies can be broken down into two subsets:  1)
those which utilize measurements of central radial velocity dispersions from
integrated spectra (c.f., Illingworth 1976), and 2) radial velocity
measurements of individual member stars (c.f., Gunn and Griffin 1979, Meylan
and Mayor 1986, Lupton \et 1987, Pryor \et 1989 and 1991 and others).

The Large Magellanic Cloud (LMC) clusters occupy a much wider range in
parameter space (i.e. age, metallicity, morphology, etc., see Olszewski
\et 1991) than their Milky Way counterparts and, hence, provide a more
complete cluster sample. To date there have been several dynamical studies
of LMC clusters including integrated spectra for several old clusters (Elson
and Freeman 1985; Dubath \et 1990; and Mateo \et 1991) and individual stellar
velocity measurements of mostly young clusters (Lupton \et 1989, Mateo \et
1991, Fischer \et 1992).

An interesting aspect of the cluster age distribution is that it does not
appear to be a continuum. There are approximately eight old (i.e. $>
10^{10}$ yrs) clusters and then a large number of clusters younger than $3
\times 10^9$ yrs, possibly indicating two major epochs of star formation
interrupted by a more quiescent period (Olszewski \et 1991).  Another
interesting feature of some LMC clusters is the existence of projected
ellipticities as large as $\epsilon \approx 0.3$ (Geisler and Hodge 1980) - a
feature not restricted to the young clusters.  For this reason, we have
embarked upon a project to dynamically study a sample of LMC clusters
covering a range of ages and ellipticities.

NGC 1978 is an intermediate age cluster ($\tau = 2 \times 10^9$ years,
Olszewski 1984 and Mould and Da Costa 1988) which also happens to be among
the most highly elliptical clusters known ($\epsilon = 0.3$, Geisler and
Hodge 1980). Three explanations for this ellipticity immediately suggest
themselves:  rotation, an anisotropic velocity dispersion tensor, or a
recent cluster-cluster merger. In this paper we will investigate the
relative likelihoods of these three scenarios and attempt to constrain the
dynamics of this largely unrelaxed object (except, perhaps, in the innermost
regions).

A discussion of the the surface photometry data and reductions will
be presented in \S 2, with the modeling results appearing in \S 3.
\S 4 contains a description of the spectroscopic observations and reductions
and in \S 5 we calculate evolutionary timescales. \S 6 details the M/L
determinations and constraints on the internal dynamics, and in \S 7 we
compare our findings with two previous studies.

\medskip
\centerline{2. SURFACE PHOTOMETRY}
\medskip

\centerline{\it 2.1 Observations and Reductions}
\medskip

In order to derive a surface brightness profile for NGC 1978, BV CCD frames
were obtained on the Las Campanas Observatory (LCO) 1.0 m telescope on 1991
January 22.  The TEK2 1024$^2$ chip was used (readout noise = 6 e$^-$, gain
= 2 e$^-$/ADU, and angular scale = 0.61$^{\prime\prime}$ pix$^{-1}$). The
integration times were 200s and 100s for the B and V frames, respectively.

There are two complications to measuring the surface brightness profile of
NGC 1978. First, the cluster is in an area densely populated with LMC field
stars, and second, elliptical apertures are necessary.

To maximize the surface brightness intensity range, it is necessary to get an
accurate estimate of the field contribution. The apparent BV color-magnitude
diagram (Fig. 1) illustrates the richness of the NGC 1978 field.  One can
see a clearly delineated red giant branch consisting of both cluster and LMC
field stars. However, there are also a similar number of blue, main sequence
LMC field stars. The presence of these bright stars complicates the
background determination and therefore it is advantageous to remove them.
This was accomplished using the profile-fitting photometry package DAOPHOT
(Stetson 1987). All stars bluer than B-V = 0.48 and all stars brighter than
V = 15.1 were removed (above and to the left of the solid line in Fig. 1).
We found that after star-subtraction the mean background brightness was only
reduced by about 10-15\% but that its uncertainty decreased by more than a
factor of ten.

Because NGC 1978 is one of the most elliptical clusters known, it requires
the use of elliptical annuli for surface photometry which in turn requires
knowledge of the mean cluster ellipticity and major axis position angle.
The many resolved stars greatly complicate attempts to derive ellipticities,
tending to skew them towards high values and yielding rapidly varying
position angles as a function of radius. Therefore, we first subtracted out
the resolved (both background and cluster) stars using DAOPHOT and then
median filtered the star-subtracted image using a 9$^{\prime\prime}$ radius
circular filter. The result was a relatively smooth light distribution with
the contribution from the bright giants largely eliminated. It was then
possible to use the ELLIPSE task in the IRAF\footnote{$^1$}{IRAF is
distributed by the National Optical Astronomy Observatories, which is
operated by the Association of Universities for Research in Astronomy, Inc.,
under contract to the National Science Foundation.} STSDAS package, which
uses the technique of Jedrzejewski (1987), to fit elliptical contours to a
smooth light distribution. We were only able to employ this technique
reliably in the range 2.5 $\le$ \Rc2 (pc) $\le$ 25 (Throughout this paper,
we assume a distance to NGC 1978 of 50 kpc and hence an angular scale of
4.12\sec pc$^{-1}$) due to crowding at smaller radii and a lack of light at
larger radii.

Table 1 lists the ellipse parameters as a function of major axis radius.
Column 1 is the radius of the projected major axis in pc, while columns 2
and 4 are the projected ellipticities and columns 3 and 5 are the major axis
position angles (PA) for the B and V frames, respectively. The tabulated
uncertainties represent the formal fitting errors. Fig. 2 shows isophotal
contour diagrams for the star-subtracted median-filtered BV frames.
Superimposed on the contours are the best-fit ellipses. The ellipses appear
to provide very reasonable models for the isophotes in the range shown.
There is no evidence for subclustering, indicating that NGC 1978 is unlikely
to be a recent merger of two clusters.  Finally, 14 stars from the Guide Star
Catalog (Lasker
\et 1990, Russell \et 1990, and Jenkner \et 1990) that appeared on the V
frame were used to accurately align the frames to the J2000.0 equinox.  The
rms residuals for the positional fit were less than 0.8 arcsec. The cluster
center is located at $\alpha$(2000) = 5:28:44.8 and $\delta$(2000) =
-66:14:09.9 .

For both frames, a mean representative ellipticity and position angle were
determined and these are shown at the bottom of Table 1.  These values agree
well with those obtained photographically by Geisler and Hodge (1980)
($\epsilon = 0.30 \pm$ 0.06 and PA = 159$^\circ \pm 7.0^\circ$) and were
used to construct the elliptical annuli for the flux measurements.  Surface
photometry was performed in a manner similar to Djorgovski (1988). The
frames were broken up into a series of concentric elliptical annuli centered
on the cluster.  The annuli were further divided into eight azimuthal
sectors. The average pixel brightness was determined for each sector in a
given annulus and the {\it median} of the eight separate measurements was
taken as the representative brightness at the area-weighted average radius
of the annulus (i.e., the mean radius of all the pixels within the annulus
which is approximately equal to the geometric mean).  The standard error of
the median of the eight sectors is equal to the standard error of the mean
multiplied by $\sqrt{\pi/2}$ and this was adopted as the photometry
uncertainty in each annulus.  Comparisons between photometry measured with
elliptical and circular apertures revealed that, while the circular
apertures did not introduce significant systematic errors, they tended to
have substantially higher sector-to-sector variations which resulted in
greater scatter in the surface photometry.

A background level (a combination of sky light and Galactic foreground and
remaining LMC field stars) was estimated from regions at large projected
distances from the cluster.  We found that the surface brightness profiles
tended to level out beyond 65 pc for the B frame and about 55 pc for the V
(both of these distances are along the major axis). By ``levelling out'' we
don't necessarily mean that the cluster light does not extend beyond this
point but simply that fluctuations in the background dominate to such an
extent that it is no longer possible to observe the profile declining in
intensity. Therefore, it was this region that was used for the background
determinations.

The cluster was reobserved on 1991 Dec 7 in order to calibrate the
photometry. Fourteen BV observations of 8 E-region standards (Menzies \et
1989) were observed on the same night covering an airmass range of 1.1 - 2.0
and a color range of 0.1 mag $\le$ B-V $\le$ 1.6 mag. The rms of the adopted
solution was less than 0.02 mag and the zero point had a similar accuracy.
E$_{B-V}$ = 0.06 mag, consistent with Olszewski (1984), was adopted. The
background-subtracted surface photometry data is presented in Table 2
[assuming M$_{V\odot} = 4.83$ and (B-V)$_\odot = 0.65$, Mihalas and Binney
1981, p. 60]. Columns 1 and 4 are the projected area-weighted radii, columns
2 and 5 are the projected major axis radii and columns 3 and 6 are the B and
V surface profiles, respectively. The background values for the B and V
frames were, respectively, $134.3 \pm 0.4$ L$_B$\sol\ pc$^{-2}$ and $148.1
\pm 0.7$ L$_V$\sol\ pc$^{-2}$. Fig. 3 is a plot of the B and V surface
brightness profiles. Also shown is a typical stellar profile which has a
FWHM approximately a factor of ten smaller than the core radius (see
\S 3); hence seeing will have a negligible effect on measurements of this
quantity (Mihalas and Binney 1981, p. 315).  We looked for a radial B-V color
gradient in the cluster but concluded that no significant effect is present
within our measurement uncertainties.

\medskip
\centerline{\it 3. KING-MICHIE MODELS FOR THE SURFACE PHOTOMETRY}
\medskip

Despite the fact that the highly elliptical shape of NGC 1978 is a clear
indication that its dynamics cannot be adequately described within the
standard framework of the King-Michie (KM) formulation (King 1966 and Michie
1963), we fitted multi-component KM models to the photometry data. The
reasons for this are: 1) These models provide us with classification
parameters adhering to a widely used scheme and thus enable potentially
fruitful comparisons to other clusters. 2) The fit provides for a reasonably
accurate means of determining the total cluster light and 3) the fit, in
conjunction with the radial velocity data (\S 4), gives a first approximation
to the cluster mass. Alternate models will be discussed in \S 6.2.

For the KM models, the stellar mass spectrum is sub-divided into mass
classes; nine were used for NGC 1978 (see Table 3). Each mass class has an
energy and angular momentum per unit mass ($E$ and $J$, respectively)
distribution function given by $$f_i(E=-0.5v^2+W,J)
\propto e^{-[J/(2v_sr_a)]^2}[e^{A_iE}-1],\eqno(\equat)$$ where $v_s$ is the
scale velocity, $W$ is the reduced gravitational potential, and r$_a$ is the
anisotropy radius beyond which stellar orbits become increasingly radial.
The $A_i$ are constants discussed below. The details are thoroughly
described in Gunn and Griffin (1979), but briefly, the shape of the density
distribution for the KM models, $\rho$(r), is determined by solving
Poisson's equation and is dependent on three parameters: the reduced central
potential $W_0$, the anisotropy radius r$_a$, and the slope of the mass
function, $x$, given by $$\phi(m) = m^{-(x+1)} ~dm~~~~~~~m \ge
0.3\hbox{M}_\odot,\eqno(\equat)$$ $$\phi(m) = m ~dm~~~~~~~~~~~~~ m <
0.3\hbox{M}_\odot.\eqno(\equat)$$ Scaling is applied in both the radial
(r$_s$) and luminosity dimensions to yield the best fit to the surface
photometry data. $A_i$ is forced to be proportional to the mean mass of
stars in the $i^{th}$ mass class, which approximates equipartition of energy
in the cluster center.

The reason that choosing the $A_i$ in the above manner does not yield true
equipartition of energy is that the lower mass stars are more affected by
the energy cut-off implicit in the model than are the higher mass stars
(Pryor \et 1986).  It turns out that the deviations in equipartition seen in
the KM models are qualitatively similar to what is calculated in theoretical
multi-mass evolutionary models. That is, they both exhibit a tendency for
the high-mass stars to have higher kinetic energies. For example, in the $x
= 0.0$ isotropic KM model described below, the lowest mass class has roughly
one-third the kinetic energy of the highest mass class at the cluster
center. In Fokker-Plank models, the cluster relaxes and equipartition begins
to occur.  Specifically, the kinetic energy of the higher mass stars
decreases while that of the lower mass stars remains fairly constant. This
results in mass segregation, with the high mass stars migrating to smaller
radii and hence an effective decoupling of the different mass classes. With
the greatly lowered rate of interactions, it becomes difficult for energy
exchange between different mass stars to occur and core equipartition cannot
be fully achieved (Inagaki and Saslaw 1985, see also Spitzer 1969 for an
analytical treatment of a two-mass system).

In order to apply the multi-mass KM models to the NGC 1978 data we adopted
stellar BV mass-luminosity relationships which were a combination of
Bergbusch \& VandenBerg (1992) in the range 0.15 $\le$ m (M$_\odot$) $\le$
0.70 and Bertelli \et (1990) in the range 0.70 $\le$ m (M$_\odot$) $\le$
1.65.  We assumed an age of $2 \times 10^9$ years and a metallicity of z =
0.004 (Mould and Da Costa 1988, and Olszewski 1984). The models of Bertelli
\et incorporate convective overshooting and mass-loss, both associated with
high mass stars, while the VandenBerg models incorporate neither of these
and should be valid in the low mass regime. The treatment of Pryor \et 1986
was adopted for remnants:  stars with initial masses of 1.65 - 4.0 M\sol\ and
4.0 - 8.0 M\sol\ become white dwarfs with masses of 0.7 M\sol\ and 1.2 M\sol,
respectively. These objects are added to the corresponding mass bins. More
massive stars, which have presumably evolved into neutron stars, are assumed
to be ejected from the cluster in agreement with the typically large
velocities observed for these objects in the field (Gunn and Griffin 1979).
Clearly, this assumption is not strictly correct as pulsars have been
detected in several galactic globular clusters.

Tables 4 and 5 (B and V, respectively) contain the fitted KM parameters for
models with parameters ranging from isotropic orbits to r$_a$ = 5 r$_s$ and
0.0 $\le$ x $\le$ 2.0.  Column 1 is the anisotropy radius, column 2 is the
mass function slope, column 3 is the reduced central potential, column 4 is
the scale radius, column 5 is the ratio of the tidal radius to the scale
radius and columns 6 and 7 are the reduced chi-squared for the fit (B has 18
and V has 17 degrees of freedom) and the probability of exceeding this value,
respectively. These probabilities are based on 1000 Monte Carlo simulations
for each parameter set, each using a surface profile generated from the best
fit model with errors drawn from the uncertainties shown in Table 2. Hence,
the probabilities are somewhat dependent on the accuracy of the photometry
uncertainties and for this reason it is safer to view them in the relative
sense. All the KM parameters are based on using surface brightness profiles
expressed as a function of area-weighted radius (from columns 1 and 4 of
Table 2). The remaining columns of the two tables will be discussed in \S
6.1.

Tables 6 and 7 contain the derived model parameters: columns 1 and 2 specify
the anisotropy radius and mass function slope, while columns 3 and 4 contain
the central luminosity density and total cluster luminosity, respectively.
Columns 7 and 8 are the central and global $population$ M/L's given by
$${M \over L_V} =
{\int^{m_u}_{m_{l}}m\phi(m)dm \over
\int^{m_u}_{m_{l}}l(m)\phi(m)dm},\eqno(\equat)$$ where $l(m)$ is the
luminosity of a star of mass $m$ given by a theoretical mass-luminosity
relationship for main-sequence and evolved stars and $m_l$ and $m_u$ are the
lower and upper mass cut-offs, respectively.  The remaining columns will be
discussed in \S 6.1.

We have plotted two of the KM models in Fig. 3; isotropic single-mass models
and isotropic $x$ = 1.0 models. The difference in the quality of the fits
results from the tendency of the multi-mass models to have shallower density
fall-offs which is not seen in the data.

The results of the KM surface brightness modeling can be summarized as
follows:  1) The parameters derived from the two surface profiles are, in
general, consistent. The B band profile, despite having smaller tabulated
uncertainties, appears to yield substantially lower $\chi^2_\nu$. Clearly,
the B profile is much less sensitive to local luminosity fluctuations caused
by cluster giants. 2) The single-mass models give the best quality fits to
the data for all values of r$_a$ with the best agreement observed for the
isotropic model.  3) For the multi-mass models there is a trend toward
poorer quality fits with steeper mass functions but the surface brightness
distributions are not sufficiently well-determined to confidently choose a
specific model. 4) The surface photometry is inadequate to unambiguously
discriminate between models with differing degrees of anisotropy. 5) For the
multi-mass models the global population M/L's are not well constrained,
varying by a factor of five.  However, the central population M/L's occupy a
much narrower range, only varying between 0.08 - 0.14 M\sol/L$_B$\sol\ and
0.10 - 0.17 M\sol/L$_V$\sol. Because the integrated cluster luminosity is
overwhelmingly attributable to the giants, the total number of giant stars
is fairly insensitive to the slope of the mass function (i.e.  one needs the
same number of giants to produce the cluster luminosity).  As a result of
the equipartition of energy, the low mass stars tend to be located at large
radii and, due to their relatively low luminosities, have little effect on
the measured surface brightness profile; the total cluster luminosity varies
by only 10\% for the various models. Hence the composition (and therefore
the M/L) of the cluster core is largely independent of the mass function
slope. As we shall see when we discuss the KM kinematic modeling in \S 6.1
the dynamical M/L's exhibit similar behavior.
\medskip
\centerline{4. RADIAL VELOCITIES}
\medskip

Spectra of 36 red giants were obtained during 1991 January 18-20 using the 4m
at Cerro Tololo Inter-American Observatory (CTIO) and during January 30
-February 1 and February 14-20 at the 2.5m at LCO. Echelle spectrographs
with 2D-Frutti detectors were employed at both telescopes.

The observation and reduction procedures for a previous run at LCO have been
discussed extensively in Welch \et 1991 and remain largely unchanged for
this data. The CTIO data were obtained and reduced in a similar manner.
Unfortunately, due to technical problems involving the dither on the CTIO
2D-Frutti, only half the available observing time was productive.
Furthermore, the spectral resolution was about 50\% lower than the LCO
spectrograph, resulting in velocities with uncertainties about 50\% larger.
The observing procedure consisted of exposures with integration times of 500
- 1500s and Th-Ar arcs approximately every 45 minutes. A representative LCO
spectrum is shown in Fig. 2 of C\^ot\'e \et (1991).  The reduction utilizes
the IRAF ECHELLE and RV packages (Tody 1986) to obtain both velocities and
velocity uncertainties. The velocity zero-point is tied to the IAU velocity
standard 33 Sex as described in Fischer \et 1992 and is believed to be
accurate to better than 1 km s$^{-1}$.

The radial velocity data are presented in Table 8. Column 1 contains the
stellar identifications, column 2 indicates the observatory, column 3 has
the projected radius, column 4 the equinox J2000.0 position angle, column 5
contains the radial velocities and column 6 contains the mean velocity for
stars with repeated measurements. Column 7 is the Heliocentric Julian Date --
2448000 for the velocity measurements. Columns 8 and 9 are V and B--V for
the stars. The velocity uncertainties returned by the RV package seem to
agree fairly well with the observed scatter in stars with repeated
measurements.  Ten stars have been measured at least twice (a total of 24
spectra), yielding $\chi^2 = 13.2$ for 14 degrees of freedom. Closer
examination of Table 8 reveals that Star 13 has two measurements which are
significantly discrepant. Further, the higher precision velocity is more
than 5 km s$^{-1}$ larger than any other cluster star. The radial velocity
implies that this star is in the LMC and, therefore, definitely a giant. The
large velocity change over about 24 hours argues against the star being a
binary and further observations are required to determine its true nature.
Alternatively, it might simply be a case of measurement error.  Regardless,
we choose not to include it in the following analysis.  Removing Star 13
reduces $\chi^2$ to 6.5 for 13 degrees of freedom. This rather low value of
$\chi^2$ is a strong indication that we have not underestimated the velocity
uncertainties and that there are no significant zero-point differences
between spectra taken on different nights or on different telescopes. Fig.
4a is a finder chart for stars 1 through 36, while Fig 4b shows the
positions of the stars relative to the cluster center along with a line
indicating the photometric minor axis.

Fig. 5 shows mean radial velocity vs. projected radius (upper panel) and
versus position angle (lower panel). The solid line is the mean velocity (\S
6), $\vave = 293.3 \pm 1.0$ km s$^{-1}$, which is consistent with the two
lower precision velocities obtained by Olzsewski
\et (1991) of 292.0 km s$^{-1}$, indicating no serious zero-point problems.
There are no obvious trends present in the data such as one might expect.
Typically, cluster velocity data will exhibit a decreasing velocity
dispersion with increasing projected radius and, if rotation is present and the
cluster
inclination favorable, a sinusoidal functional dependence on position angle
(Fischer \et 1992). The reason we do not observe either of these phenomena
may simply be due to the sparseness of the data and the large values of the
velocity uncertainties relative to both the velocity dispersion and the
rotation amplitude.

\medskip
\centerline{5. EVOLUTIONARY TIMESCALES}
\medskip

Implicit in the adoption of ``mass segregation'' models is the assumption
that there has been sufficient time for energy exchange between different
mass classes to occur. Two important relaxation timescales based on energy
exchange through distant two-body encounters are the central relaxation time
 $$t_{r0} = (1.55 \times
10^7
\hbox{yr})\left({r_s
\over \hbox{pc}}\right)^2\left({v_s \over \hbox{km s}^{-1}}\right)
\left({M_\odot
\over \left< m \right>} \right)\left[\log(0.5 M / \left< m
\right>)\right]^{-1} = 2.9 - 7.9 \times 10^8 \hbox{yr},\eqno(\equat)$$
(Lightman and
Shapiro 1978) and the half mass relaxation time $$t_{rh}=(8.92 \times 10^8
\hbox{yr})\left({M
\over 10^6 M_\odot}\right)^{1/2}\left({r_h \over \hbox{pc}
}\right)^{3/2}\left({M_\odot \over
\left< m \right>}\right)\left[\log(0.4M/\left< m \right>)\right]^{-1} = 6.0
- 16.0 \times 10^9 \hbox{yr},\eqno(\equat)$$ (Spitzer and Hart 1971), where
the numerical values are for mass functions slopes ranging from 0.0 $\le x
\le$ 2.0.  As was mentioned earlier, the best age estimate for NGC 1978 is $2
\times 10^9$ yrs which is several times greater than the central relaxation
time, but significantly younger than the half-mass relaxation time. We
conclude, therefore, that there are significant portions of the cluster
which have not had sufficient time to relax and it is important to keep this
in mind when interpreting the results of the KM model analysis.

\medskip
\centerline{6. MASS DETERMINATIONS}
\medskip
\centerline{\it 6.1 King-Michie Models}
\medskip

The mass of a multi-mass KM model is given by $$M = {9r_sv_s^2 \over 4\pi G}
\int{{\rho \over \rho_0}r^2dr}\eqno(\equat)$$ Illingworth (1976), where r$_s$
is given
in Tables 4 and 5, and $v_s$ is the scale velocity. The run of
$\sigma^2_{r,i}(r)$ and $\sigma^2_{t,i}(r)$ are determined from
$$\sigma_{(r,t),i}^2(r) = {\int_{|\sigma_i|
\le W(r)} f_i(\sigma_i,W)\sigma_k^2d^3\vec \sigma_i \over \int_{|\sigma_i|
\le W(r)} f_i(\sigma_i,W)d^3\vec\sigma_i},\eqno(\equat)$$ where $W$ is the
reduced potential (W = 0 at the tidal radius), $\sigma_k =
\sigma_i$cos$\theta$ or $\sigma_i$sin$\theta$ for $\sigma_{r,i}$ or
$\sigma_{t,i}$, respectively, and the $i$ subscript refers to the $i^{th}$
mass class. Comparisons were made between the observed velocities and scaled
model velocity dispersions projected along the line of sight,
$$\sigma_{p,i}^2(R) = {2 \over
\mu_i (R)}
\int^\infty_ R{\rho_{i}(r)[(r^2- R^2)\sigma_{r,i}^2(r)+ R^2\sigma_{t,i}
^2(r)]dr
\over r(r^2- R^2)^{1/2}},\eqno(\equat)$$ (Binney and Tremaine 1987, p. 208),
where $\mu_i$ is the surface density of the $i^{th}$ mass class. The optimal
scaling was derived using the maximum likelihood technique outlined in Gunn
and Griffin (1979). Simply put, the probability density function for
v$_{x,k}$, an observed stellar velocity, is a Gaussian with standard
deviation equal to the model dispersion added in quadrature to the velocity
uncertainty:  $$P_k \sim {1 \over
\sqrt{\sigma_{err,k}^2+v_s^2\sigma_{p,k,i}}} e^{-(v_{x,k} - \vave)^2/2(v_s^2
\sigma_{p,k,i}^2 + \sigma_{err,k}^2)}.\eqno(\equat)$$ This function is
minimized with respect to $v_s$ and $\vave$ resulting in two equations which
can be solved simultaneously for the most probable values of the two
parameters.

The values of $v_s$ thus obtained are displayed in column 8 of Tables 4 and
5. The corresponding dynamical masses and M/L's are in columns 9 and 10,
respectively of Tables 6 and 7. Monte-Carlo orbit simulations were used to
determine the uncertainties in the fitted and derived parameters and to
search for possible systematic effects.  We started with the known projected
radii (\Rc2$_k$) of the program stars. The true radii are in the range
\Rc2$_k$ $\le$ r $\le$ r$_{t}$. If x is the displacement from the mean
cluster position along the line-of-sight such that r =
$\sqrt{R_k^2+{\hbox{x}}^2}$, then the probability that a star is at x is
$p$(x), where $$p(x)
\sim
\rho_{i}(\sqrt{R_k^2+x^2}).\eqno(\equat)$$ Three-dimensional positions,
along with corresponding model-dependent radial and tangential velocities
were drawn at random from their respective probability distributions. The
velocity component along the line-of-sight was then determined, and an error
term, drawn from a Gaussian distribution with standard deviation equal to the
velocity uncertainty, as tabulated in Table 8, was added. This process was
repeated, producing 10000 sets of data each with a given mass, r$_a$, x and
the same projected positions and velocity measurement errors as the original
data set. Finally the maximum likelihood technique was applied to each of
the artificial data sets and the results compared to the input values for
the models. From this we noticed that the maximum likelihood method resulted
in scale velocities that were biased systematically too low by approximately
4\% (the values for v$_s$ in Tables 4 and 5 have already been corrected
for this effect).

A goodness-of-fit statistic $$\zeta^2 = \sum{{(v_{x,k} - \vave)^2} \over
(v_s^2 \sigma_{p,k}^2 + v_{err,k}^2)}\eqno(\equat)$$ was generated for each
model and is shown in column 9 of Tables 4 and 5 (33 degrees of freedom). The
distribution of this statistic can be extracted from the Monte Carlo
simulations and column 10 shows the probability of exceeding the observed
$\zeta^2$ assuming that the cluster velocities are specified by the model
parameters indicated and have the uncertainties tabulated in Table 8. The
greater this probability the higher the likelihood that the cluster
velocities are drawn from the specified distribution.

The results of the KM kinematic modeling can be summarized as follows: 1)
The results for the two different bandpasses are consistent. 2) As can be
seen from the P($> \zeta^2$), the radial velocity data are too sparse to
provide a means of discriminating between the different sets of dynamical
parameters with any confidence. However, the multi-mass models are in
marginally better agreement with the kinematic data than the single-mass
models.  3) A similar trend is seen in the dynamical M/L's as was seen for
the population M/L's: the global M/L is very poorly determined
while the central M/L is much more tightly constrained. As $x$ is increased,
the number of low-mass stars at large radii is increased. The resultant
change in the gravitational potential at points where we have measured
stellar velocities is minor, and therefore there is very little change in
v$_s$. As was mentioned in \S 3, there is also very little change in the
central luminosity density and, therefore, a relatively model-independent
central M/L. 4) The best agreement between population and dynamical M/L's is
seen for the cases $x$ = 0.0 for the B band [(M/L)$_0 = 0.14 \pm$ 0.06
M\sol/L$_B$\sol\ and M/L = 0.34 $\pm$ 0.15 M\sol/L$_B$\sol] and $x$ = 0.5
for the V [(M/L)$_0 = 0.13
\pm$ 0.06 M\sol/L$_V$\sol\ and M/L = 0.40 $\pm$ 0.15 M\sol/L$_V$\sol],
independent of r$_a$. However, there is agreement at fairly high confidence
levels for all values of $x$. 5) Because the velocity uncertainties are
about 70\% of the derived velocity dispersion, the cluster mass estimates are
somewhat dependent upon the accuracy of these uncertainties. As discussed
earlier, the $\chi^2$ for the 10 stars with multiple measurements came out
lower than average indicating that the uncertainties are probably not
underestimated.  If, however, they are overestimated then one would get an
underestimate for both the velocity dispersion and the mass. The worst
possible (not to mention unreasonable) case is that the uncertainties are
actually zero. In this case, we get a slightly less than 30\% increase in
the velocity dispersion and about a 60\% increase in the cluster mass and
M/L.

\medskip
\centerline{\it 6.2 Oblate Spheroids}
\medskip

In order to determine the cause of flattening we have decided to model NGC
1978 as both a rotating and non-rotating (i.e. anisotropically supported)
oblate spheroid. Because of the sparseness of the kinematic data, the goal
will be to construct models with {\it extreme} sets of parameters and see
which provide the greatest consistency with the data. This will also yield a
mass range for the cluster.

For an axisymmetric system, the relevant Jeans' equations (velocity moments
of the collisionless Boltzmann equation) in cylindrical coordinates are:
$${\partial(\rho \sigma_{\hbox{\fsmall R}}^2) \over \partial \hbox{R}} +
{\partial(\rho\sigma_{\hbox{\fsmall R}z}) \over \partial z} +
\rho\left({\sigma_{\hbox{\fsmall R}}^2 - \sigma_\phi^2 - v_\phi^2 \over
\hbox{R}}\right) + \rho{\partial\Phi \over \partial \hbox{R}} =
0,\eqno(\equat)$$ and $${\partial(\rho\sigma_{\hbox{\fsmall R}z}) \over
\partial \hbox{R}} + {\partial(\rho\sigma_z^2) \over \partial z} +
{\rho\sigma_{{\hbox{\fsmall R}}z}
\over \hbox{R}} + \rho{\partial\Phi \over \partial z} = 0,\eqno(\equat)$$
where (R,$\phi$,z) are the cylindrical coordinate axes, the
($\sigma_{\hbox{\fsmall R}}, \sigma_\phi,
\sigma_z$) are the corresponding velocity dispersions, and $v_\phi$ is the
rotation velocity. $\Phi$ is the gravitational potential.

Both the rotating and non-rotating models which were used have velocity
ellipsoids aligned with the cylindrical coordinate axes (i.e.
$\sigma_{\hbox{\fsmall R}z}=0$) and, as well, both have $$\sigma_\phi =
{\sigma_{\hbox{\fsmall R}}
\over \sqrt{1 + (\hbox{R}/\hbox{R}_a)^2}},\eqno(\equat)$$ where $\hbox{R}_a$ is
varied
from 5 pc to $\infty$. The rotating models also have the condition
$$\sigma_{\hbox{\fsmall R}} = \sigma_z,\eqno(\equat)$$ implying $$v_\phi^2 =
\hbox{R}{\partial\Phi \over
\partial \hbox{R}} + {\hbox{R} \over \rho}{\partial \over \partial
\hbox{R}}\int_z^\infty{\rho
{\partial\Phi \over \partial z}dz} + {1 \over
\rho}\left[1 - {1 \over
\sqrt{1 + (\hbox{R}/\hbox{R}_a)^2}}\right]\int_z^\infty{\rho{\partial\Phi \over
\partial
z}dz}.\eqno(\equat)$$

The models are constructed by assuming that the mass distribution is
equivalent to the deprojected light distribution (constant M/L).  Equations
14 and 17 can then be solved directly to obtain $\sigma_z$, and $v_\phi$.
Once $v_\phi$ is known it can be substituted into equation 13 which, in turn,
can be solved for $\sigma_{\hbox{\fsmall R}}$ and $\sigma_\phi$. This is
outlined in Binney and Tremaine (1987) for the $\sigma_{\hbox{\fsmall R}} =
\sigma_\phi = \sigma_z$ case.

Fig. 6 shows isovelocity maps for the models with $\hbox{R}_a =
\infty$. The top panel is $\sigma_{\hbox{\fsmall R}}$ for the rotating model
and the middle panel is the corresponding $v_\phi$. The bottom panel is
$\sigma_{\hbox{\fsmall R}}$ for the non-rotating model. Unfortunately, it is
impossible to determine the inclination of NGC 1978 so we assume that it is
90$^\circ$.  We believe this to be reasonable as it is already among the
most elliptical clusters known and, of course, if it is inclined then it is
intrinsically even more elliptical.

Once the models have been generated, it is simply a matter of scaling them
using a similar maximum likelihood method to that employed for the KM
models. Table 9 shows the results of this exercise; column 1 is
$\hbox{R}_a$, while columns 2, 3, 4 and 5 are, respectively, the total mass,
the goodness-of-fit parameter $\zeta^2$, the probability of exceeding it (as
described in
\S 6.1) and the cluster M/L$_V$ for the non-rotating case. Similarly, columns
6, 7, 8, and 9 represent the rotating case.

There are four points worth noting: 1) The distribution of $\zeta^2$ is much
broader for the rotating models. Fig. 7 shows histograms of $\zeta^2$ based
on sets of 10000 Monte Carlo simulations for two $\hbox{R}_a = \infty$
models. The solid line is the non-rotating model while the dashed line is
the rotating model. It is because of this broader distribution that we cannot
completely reject the rotating models.  2) The rotating models produce
higher masses than the non-rotating models, opposite to what is seen in NGC
1866 (Fischer \et 1992), a cluster which has rotation detected at the 97\%
confidence level.  In fact, higher masses should result from the application
of a rotating model to a non-rotating system since the removal of a
non-existent rotation field causes an increase in the apparent velocity
dispersion. 3) The oblate spheroid masses are consistent with the spherical
single-mass KM model masses derived in \S 6.1. Even with $\epsilon
\approx 0.3$ a spherical approximation is excellent. 4) The non-rotating
(anisotropic velocity dispersion) models appear to represent a better fit to
the data, although it is impossible to exclude the rotating models,
especially those with higher values of $\hbox{R}_a$. This trend to higher
P($>\zeta^2$) with increasing $\hbox{R}_a$ is seen in the non-rotating case
as well. The precision and size of our radial velocity data set is
insufficient to strongly constrain the flattening mechanism.

There is very little evidence to support the hypothesis of a recent
cluster-cluster merger. The cluster light distribution is very smooth, the
isophotes agree well with ellipses with no evidence for subclustering, and
there is no sign of tidal interaction.  Furthermore, except in very
restricted cases, a recent merger would tend to give the cluster a net
rotation for which there is no strong indication.

Because these are single mass models (i.e. the mass scales as the
luminosity) they do not explicitly yield information about the mass function.
However, it is possible to test which mass function is consistent with the
dynamical M/L$_V$. We maintain the same assumptions about the form of the
mass function (i.e. a power-law with a flattening at 0.3 M$_\odot$), the
mass-luminosity relationship and the mass of the stellar remnants that were
used for the KM modeling.

The population M/L$_V$ is given by equation 4.  We conclude that it is very
difficult to reconcile the low M/L$_V$'s with the adopted form of the mass
function without invoking a high low-mass cut-off. With the KM low-mass
cut-off (0.15 M$_\odot$), the lowest population M/L$_V$ that can be achieved
is M/L$_V$ = 0.40 M\sol/L$_V$\sol\ at $x \approx 0.2$. It is necessary to
raise the low-mass cut-off to 0.8 M\sol\ (with $x = 1.6$) in order to get a
population M/L$_V$ of about 0.2 M\sol/L$_V$\sol. This seems unreasonably
high for a low-mass cut-off and leads us to believe that either our adopted
mass function is an oversimplification or the assumption that mass follows
light is unreasonable.  Certainly, we see from the multi-mass KM models that
it is possible to get excellent agreement between population and dynamical
M/L$_V$'s supporting the latter supposition.

\medskip
\centerline{7. COMPARISON WITH PREVIOUS RESULTS}
\medskip

There have been two previous kinematic studies of NGC 1978, and we consider
these in turn.  Meylan \et (1991) find a central velocity dispersion of 5.8
$\pm 1.2$ km s$^{-1}$ based upon an integrated spectrum of the cluster
center (they use a region that does not overlap with our radial velocity
data set). Seitzer (1991) finds a dispersion of 3.7 $\pm$ 0.8 km s$^{-1}$
based upon 8 stellar velocities within two core radii of the cluster center.
Seitzer further claims that the central velocity dispersion is a
model-dependent 10 - 40\% higher than this number. Our central velocity
dispersion is also model-dependent and is about $\sigma_0 = 2.2 \pm$ 0.5
km s$^{-1}$.

Our result disagrees with Meylan \et at the 2.8$\sigma$ level and, even if we
assume that our radial velocity uncertainties are zero, we get an upper limit
of $\sigma_0 = 2.8 \pm 0.6$ km s$^{-1}$, still more than 2.2$\sigma$ lower.
Perhaps their result is affected by a large rotation velocity at the cluster
center, but it is not possible to rule out a central mass density cusp. If
so, the surface photometry does not reveal an accompanying luminosity
density cusp. The disagreement with Seitzer is at approximately the
2$\sigma$ level and it is not possible to definitively resolve this
discrepancy without knowledge of which stars he measured.

We conclude by stating that when measuring a velocity dispersion the sources
of error one encounters, such as binaries, field stars, slit errors,
zero-point drift, etc. will tend to bias the result too high. One possible
source of error than can cause the opposite effect occurs when two (or more)
stars fall in the slit. This will tend to give the mean velocity for the two
objects, which will, on average, be lower than the residual velocities of the
individual stars. We feel that this has not been a problem with our sample of
giants which are relatively isolated and significantly brighter than the
underlying cluster light.
\medskip
\centerline{8. CONCLUSIONS}
\medskip

In this paper we have examined the internal dynamics of the elliptical LMC
cluster NGC 1978 using BV CCD images and echelle spectra of 35 giants.
Projected radii for the giants range from $1.4 \le$ R (pc) $\le 20.0$ and the
mean estimated stellar velocity uncertainty is $\sigma_{err} \approx 1.6$ km
s$^{-1}$.  The mean cluster velocity is $\vave = 293.3 \pm 1.0$ km
s$^{-1}$.

\item {1)} BV luminosity profiles were constructed out to projected radii of
R $>$ 100 pc.  Despite the large ellipticity, single and multi-mass
King-Michie models with 5.0 $\le$ r$_a$ (r$_s$) $\le \infty$ and 0.0 $\le x
\le 2.0$ were applied to the data. The single-mass models provided better
agreement with the surface photometry which is perhaps not surprising since
NGC 1978 is considerably younger than its half-mass two-body relaxation time.
Among multi-mass models, there is (slightly) better agreement seen for the
models with shallow mass functions.  The total cluster luminosity is model
dependent; L$_B = 3.1 - 3.7 \pm 0.2 \times 10^5$ L$_B$\sol\ and L$_V = 3.0 -
3.5 \pm 0.2
\times 10^5$ L$_V$\sol.

\item{2)} The single mass KM models yielded M/L = 0.20 $\pm$ 0.08
M\sol/L\sol. For the multi-mass KM models, we found that while the central
M/L's were relatively tightly constrained to be around (M/L)$_0$ = 0.13
$\pm$ 0.06 M\sol/L\sol\ the global M/L's ranged over more than a factor of
five (i.e., M/L = 0.3 - 1.5 M\sol/L\sol). The best agreement between the
population and dynamical M/L's is seen for the cases $x$ = 0.0 for the B
band [(M/L$_B$)$_0 = 0.14 \pm$ 0.06 M\sol/L$_B$\sol\ and M/L$_B$ = 0.34
$\pm$ 0.15 M\sol/L$_B$\sol] and $x$ = 0.5 for the V [(M/L)$_0 = 0.13
\pm$ 0.06 M\sol/L$_V$\sol\ and M/L = 0.40 $\pm$ 0.15 M\sol/L$_V$\sol],
independent of r$_a$. The kinematic data were too sparse to place strong
constraints on dynamical parameters such as the anisotropy radius or the
mass function.

\item{3)} Non-rotating single-mass oblate spheroid models produced M/L's
consistent with the single-mass KM models while the rotating models had
marginally higher M/L's. We found that the non-rotating model was in better
agreement with the kinematic data but that it was impossible to completely
rule out the rotating models. As well, there is very little morphological
evidence for a merger; the light distribution is quite smooth, the isophotes
are very elliptical (i.e. no subclustering) and there is no sign of tidal
interaction.

\item{4)} In order to get consistency between the single-mass dynamical M/L
and a simple power-law mass function requires an unusually high low-mass
cut-off. A more probable solution invokes the multi-mass models or perhaps a
more complex form for the mass function.

\bigskip
\centerline{ACKNOWLEDGEMENTS}
\medskip

P.F. would like to knowledge the Natural Sciences and Engineering Research
Council (NSERC) for a post-graduate fellowship and operating grant support
to D.L.W. This work was undertaken while D.L.W. was an NSERC University
Research Fellow. Partial support for this work was provided by NASA through
grant \# HF-1007.01-90A awarded by the Space Telescope Science Institute
which is operated by the Association of Universities for Research in
Astronomy, Inc. for NASA under contract NAS5-26555. We would like to thank
Dr. S. Heathcote for setting up and trouble-shooting the 2D-Frutti at CTIO.
Also thanks to Dr. M.  Roth (and indirectly to Saddam Hussein) for kindly
giving us extra observing time at LCO and to Dr. M. Merrifield for some
advice regarding the oblate spheroid models. Finally, we thank the referee,
Dr. C. Pryor for the prompt and very thorough job he did which significantly
tightened up the paper.

\bigskip

Reprints of this paper are available through mail or anonymous FTP. Contact
Fischer@Crocus.Physics.McMaster.CA for information.

\vfill\eject
\def\ref#1#2#3#4{#1, {\it #2}, {#3}, #4.
\smallskip}
\def\apj{ApJ}
\def\apjs{ApJS}
\def\aj{AJ}

\def\aa{A\&{A}}
\def\aas{A\&{A}S}
\def\pasp{PASP}
\def\mnras{MNRAS}
\def\fullref#1{#1. \smallskip}

\centerline{REFERENCES}
\medskip

{\everypar={\parindent=0pt\parskip=0pt\hangindent=20pt\hangafter=1}

\fullref{Bergbusch, P. and Vandenberg, D. A., 1992, \apjs, in press}

\ref{Bertelli, G., Betto, R., Bressan A., Chiosi C., Nasi, E., and Vallenari A.
1990}{\aas}{85}{845}

\fullref{Binney J., and Tremaine, S. 1987, in {\it Galactic Dynamics},
(Princeton: Princeton University Press)}

\ref{C\^ot\'e P., Welch, D. L., Mateo M., Fischer P., and Madore, B. F. 1991}
{\aj}{101}{1681}

\fullref{Djorgovski S. 1988, in {\it The Harlow-Shapley Symposium on
Globular Cluster Systems in Galaxies (IAU Symposium No. 126)}, eds. J. E.
Grindlay and A. G. Davis Philip, (Dordrecht: Reidel), p. 333}

\fullref{Dubath, P., Meylan, G., Mayor, M., and Magain, P. 1990, STSI Preprint
No. 443}

\ref{Elson, R. A. W., Freeman, K. C. 1985}{\apj}{288}{521}

\ref{Fischer, P, Welch, D. L., C\^ot\'e, P., Mateo, M., and Madore, B. F.
1992}{\aj}{103}{857}

\ref{Geisler, D., and Hodge, P. 1980}{\apj}{242}{66}

\ref{Gunn, J. E. and Griffin, R. F. 1979}{\aj}{84}{752}

\ref{Jedrzejewski, 1987}{\mnras}{226}{747}

\ref{Jenkner, H., Lasker, B. M., Sturch, C. R., McLean, B. J., Shara, M. M.,
and Russell, J. L. 1990}{\aj}{99}{2019}

\ref{Illingworth, G. D. 1976}{\apj}{204}{73}

\ref{Inagaki, S. and Saslaw, W. C. 1985}{\apj}{292}{339}

\ref{King, I. R. 1966}{\aj}{71}{64}

\ref{Lasker, B. M., Sturch, C. R., McLean, B. J., Russell, J. L., Jenkner, H.,
and Shara, M. M. 1990}{\aj}{99}{2019}

\ref{Lightman, A. P. and Shapiro, S. L. 1978}{Rev. Mod. Phys.}{50}{437}

\ref{Lupton, R. H., Gunn, J. E. and Griffin, R. F. 1987}{\aj}{93}{1114}

\ref{Lupton, R. H., Fall, S. M., Freeman, K. C., and Elson, R. A. W.
1989}{\apj}{347}{201}

\fullref{Mateo, M., Welch, D. L., and Fischer, P. 1991 in {\it IAU
Symposium 148, The Magellanic Clouds,} (Dordrecht: Reidel), p. 191}

\ref{Michie, R. W. 1963}{\mnras}{126}{499}

\fullref{Menzies, J. W., Cousins, A. W. J., Banfield, R. M., and Laing, J.
D. 1989, South African Astron. Obs. Circulars, 13, 1}

\fullref{Meylan, G., Dubath, P., and Mayor, M. 1991 in {\it IAU Symposium
148: The Magellanic Clouds}, p. 211, eds. R. Haynes and D. Milne, (Kluwer:
Dordrecht)}

\ref{Meylan, G. and Mayor, M. 1986}{\aa}{166}{122}

\fullref{Mihalas, D., and Binney, J. 1981, in {\it Galactic Astronomy Structure
and Kinematics}, (W. H. Freeman and Company: New York)}

\fullref{Mould, J. R., and Da Costa, G. S. 1988, in {\it Progress and
Opportunities in Southern Hemisphere Optical Astronomy}, ASP Conference
Series Vol. 1, eds. V.M. Blanco and M.M. Phillips. p. 197}

\ref{Olszewski, E. W. 1984}{\apj}{284}{108}

\ref{Olszewski, E. W., Schommer, R. A., Suntzeff, N. B., and Harris, H. C.
1991}{\aj}{101}{515}

\ref{Pryor, C., McClure, R. D., Fletcher, J. M., and Hartwick, F. D. A., and
Kormendy, J. 1986}{\aj}{91}{546}

\ref{Pryor, C., McClure, R. D., Fletcher, J. M., and Hesser, J. E. 1989}{\aj}
{98}{596}

\ref{Pryor, C., McClure, R. D., Fletcher, J. M., and Hesser, J. E. 1991}{\aj}
{102}{1026}

\ref{ Russell, J. L., Lasker, B. M., McLean, B. J., Sturch, C. R., and Jenkner,
H. 1990}{\aj}{99}{2059}

\fullref{Seitzer, P. 1991 in {\it IAU Symposium 148: The Magellanic Clouds},
p. 213, eds. R. Haynes and D. Milne, (Kluwer:  Dordrecht)}

\ref{Spitzer, L. and Hart, M. H. 1971}{\apj}{164}{399}

\ref{Stetson, P. B. 1987}{\pasp}{99}{191}

\fullref{Tody, D. 1986 in {\it The IRAF Data Reduction and Analysis System,
Instrumentation in Astronomy VI}, ed. D. L. Crawford, Proc. SPIE,
p. 713}

\ref{Welch, D. L., Mateo, M., C\^ot\'e P., Fischer, P., and Madore, B. F.
1991}{\aj}{101}{490}
}
\vfill\eject

\centerline{FIGURE CAPTIONS}

\noindent Fig. \fig -- Apparent BV color-magnitude diagram for NGC 1978.
Stars above and to the left of the solid line were subtracted prior to
surface brightness measurements.

\noindent Fig. \fig -- Plot of elliptical contours fit to star-subtracted and
median-filtered BV CCD images. The solid lines are the isophotes and the
dashed lines are the best fit ellipses.  The elliptical parameters are
tabulated in Table 1.

\noindent Fig. \fig -- CCD BV luminosity profiles for NGC 1978.  The solid
lines are the single-mass isotropic KM models, while the short-dashed lines
are the multi-mass $x = 1.0$ isotropic models. The long-dashed lines are
typical stellar profiles (FWHM approximately one tenth the cluster core
radius).

\noindent Fig. \fig -- A finder chart for stars with measured radial
velocities. North is at the top, and east is to the left.

\noindent Fig. 4b -- Positions with respect to the cluster center for stars
with measured radial velocities. The straight line is the photometric minor
axis.

\noindent Fig. \fig -- Plot of observed radial velocity versus position
angle (lower panel) and projected radius (upper panel). The solid lines
indicate the cluster mean velocity.

\noindent Fig. \fig -- Isovelocity maps for oblate spheroid models.  The top
panel is $\sigma_{\hbox{\fsmall R}}$ for the rotating model and the middle
panel is the corresponding $v_\phi$. The bottom panel is
$\sigma_{\hbox{\fsmall R}}$ for the non-rotating model.

\noindent Fig. \fig -- Histograms of $\zeta^2$ based on sets of 10000 Monte
Carlo simulations for two R$_a = \infty$ oblate spheroid models. The solid
line is the non-rotating model while the dashed line is the rotating model.

\vfill\supereject

\baselineskip=13pt
\pageinsert
\tabhl{\taone}{Ellipticity and Position Angle}
\span\tcolil{9}{4} & \span\tcol{.3} & \span\tcoll{.3}{9}{4} && \span\tcol{.3}
\cr
\topperl

 & \multispan4{\hfil \hskip 0.01truecm B \hfil \vvbs{9}{4} \hfil V \hfil} \cr
\tablerule
\hfil a & $\epsilon$ & PA & $\epsilon$ & PA \cr
\hfil (pc) & &  ($^\circ$) & & ($^\circ$) \cr
\tablerule

{}~2.9 & $0.33 \pm 0.01$ & $164.0 \pm 2.0$ & $0.29 \pm 0.01$ & $161.0 \pm 2.0$
\cr
{}~5.2 & $0.33 \pm 0.02$ & $147.0 \pm 2.0$ & $0.28 \pm 0.02$ & $158.0 \pm 3.0$
\cr
{}~7.6 & $0.25 \pm 0.03$ & $151.0 \pm 4.0$ & $0.20 \pm 0.03$ & $154.0 \pm 4.0$
\cr
10.1 & $0.28 \pm 0.02$ & $148.0 \pm 2.0$ & $0.25 \pm 0.02$ & $148.0 \pm 2.0$
\cr
13.4 & $0.31 \pm 0.04$ & $146.0 \pm 4.0$ & $0.30 \pm 0.02$ & $148.0 \pm 3.0$
\cr
17.8 & $0.31 \pm 0.04$ & $146.0 \pm 4.0$ & $0.33 \pm 0.03$ & $148.0 \pm 3.0$
\cr
21.6 & $0.34 \pm 0.03$ & $147.0 \pm 3.0$ & $0.34 \pm 0.03$ & $147.0 \pm 3.0$
\cr
\tablerule
Mean & $0.32 \pm 0.02$ & $151.0 \pm 7.0$ & $0.29 \pm 0.03$ & $153.0 \pm 6.0$
\cr
\spl
\endinsert
\vfill\dosupereject

\pageinsert
\tabhl{\taone}{Surface Photometry}
\span\tcoli & \span\tcol{.3} & \span\tcoll{.3}{9}{4} && \span\tcol{.3} \cr
\topperl

\hfil \Rc2 & a & L$_B$ & \Rc2 & a & L$_V$ \cr
\hfil (pc) & (pc) & (L$_B$\sol\ pc$^{-2}$) & (pc) & (pc) & (L$_V$\sol\
pc$^{-2}$) \cr
\tablerule

     ~0.4 & ~0.5 & $1886.0 \pm 140.0$ & ~0.4 & ~0.5 & $2406.0 \pm  305.0$ \cr
     ~0.6 & ~0.7 & $1668.0 \pm 137.0$ & ~0.6 & ~0.7 & $2142.0 \pm  199.0$ \cr
     ~0.8 & ~1.0 & $1697.0 \pm ~92.0$ & ~0.8 & ~0.9 & $1813.0 \pm  141.0$ \cr
     ~1.0 & ~1.2 & $1599.0 \pm ~88.0$ & ~1.0 & ~1.2 & $1678.0 \pm  ~88.0$ \cr
     ~1.3 & ~1.5 & $1479.0 \pm 107.0$ & ~1.3 & ~1.4 & $1469.0 \pm  189.0$ \cr
     ~1.6 & ~1.9 & $1429.0 \pm 121.0$ & ~1.6 & ~1.8 & $1674.0 \pm  146.0$ \cr
     ~2.0 & ~2.4 & $1478.0 \pm ~72.0$ & ~2.0 & ~2.3 & $1806.0 \pm  170.0$ \cr
     ~2.6 & ~3.0 & $1195.0 \pm ~58.0$ & ~2.5 & ~2.9 & $1266.0 \pm  ~77.0$ \cr
     ~3.2 & ~3.8 & $1018.0 \pm ~38.0$ & ~3.2 & ~3.6 & $1049.0 \pm  ~38.0$ \cr
     ~4.1 & ~4.8 & $~829.0 \pm ~47.0$ & ~4.0 & ~4.6 & $~910.0 \pm  ~73.0$ \cr
     ~5.1 & ~6.0 & $~687.0 \pm ~24.0$ & ~5.0 & ~5.8 & $~755.0 \pm  ~41.0$ \cr
     ~6.5 & ~7.6 & $~496.0 \pm ~22.0$ & ~6.3 & ~7.3 & $~507.0 \pm  ~30.0$ \cr
     ~8.2 & ~9.5 & $~320.0 \pm ~11.0$ & ~7.9 & ~9.1 & $~324.0 \pm  ~27.0$ \cr
     10.3 & 12.0 & $~201.0 \pm ~~8.2$ & 10.0 & 11.5 & $~213.0 \pm  ~15.0$ \cr
     12.9 & 15.1 & $~134.0 \pm ~~9.6$ & 12.6 & 14.5 & $~136.0 \pm  ~12.0$ \cr
     16.3 & 19.0 & $~~73.0 \pm ~~2.5$ & 15.8 & 18.2 & $~~78.0 \pm  ~~7.4$ \cr
     20.5 & 24.0 & $~~35.0 \pm ~~5.4$ & 20.0 & 23.0 & $~~38.0 \pm  ~~7.7$ \cr
     25.8 & 30.2 & $~~15.0 \pm ~~3.1$ & 25.1 & 28.9 & $~~15.0 \pm  ~~3.7$ \cr
     32.4 & 38.0 & $~~~8.6 \pm ~~3.7$ & 31.6 & 36.4 & $~~~7.2 \pm  ~~3.3$ \cr
     40.9 & 47.8 & $~~~3.5 \pm ~~2.3$ & 39.8 & 45.8 & $~~~3.6 \pm  ~~1.6$ \cr
     51.4 & 60.2 & $~~~0.2 \pm ~~1.9$ \cr
\spl
\endinsert
\vfill\dosupereject

\pageinsert
\tabh{\taone}{Mass Bins}
\span\tcoli && \span\tcol{.3} \cr
\topper

\hfil Bin & m$_{min}$ & m$_{max}$ \cr
\hfil  & (M\sol) &  (M\sol) \cr
\spacer

1 & 0.16 & 0.30 \cr
2 & 0.30 & 0.45 \cr
3 & 0.45 & 0.60 \cr
4 & 0.60 & 0.75 \cr
5 & 0.75 & 0.90 \cr
6 & 0.90 & 1.05 \cr
7 & 1.05 & 1.20 \cr
8 & 1.20 & 1.43 \cr
9 & 1.43 & 1.65 \cr
\sp
\endinsert
\vfill\dosupereject

\pageinsert
{
\tabhl{\taone}{King-Michie - B Band Fitted Parameters}
\span\tcoli & \span\tcoll{.15}{9}{4} & \span\tcol{.15} & \span\tcol{.15} &
\span\tcol{.15} & \span\tcol{.15}
& \span\tcoll{.15}{9}{4} && \span\tcol{.15} \cr
\topperl

 & & & \multispan3 Photometry & & \multispan3 Velocities \cr
\tablerule
\hfil r${_a}$ & x & W$_0$ & r$_s$ & c & $\chi_\nu^2$ & P($> \chi_\nu^2$) &
v$_s$ & $\zeta^2$ & P($> \zeta^2$) \cr
\hfil (r$_s$) & &  & (pc) & & ($\nu=18$) & & (km s$^{-1}$) & & \cr
\tablerule

ISO &     & $~5.9 \pm 0.2$ &$4.6 \pm 0.2$ & $~17.0 \pm ~2.0$ & 0.75 & 0.74 &
$2.33 \pm 0.5$ & 33.79 & 0.71 \cr
{}~10 &     & $~5.9 \pm 0.2$ &$4.6 \pm 0.2$ & $~18.0 \pm ~2.0$ & 0.78 & 0.70 &
$2.34 \pm 0.5$ & 33.85 & 0.70 \cr
{}~~5 &     & $~5.8 \pm 0.2$ &$4.8 \pm 0.2$ & $~22.0 \pm ~4.0$ & 0.90 & 0.53 &
$2.38 \pm 0.5$ & 33.99 & 0.67 \cr
&&&&&&&&& \cr
ISO & 0.0 & $~6.9 \pm 0.4$ &$5.6 \pm 0.2$ & $~21.0 \pm ~4.0$ & 1.21 & 0.20 &
$2.31 \pm 0.5$ & 33.00 & 0.84 \cr
{}~10 & 0.0 & $~6.9 \pm 0.4$ &$5.6 \pm 0.2$ & $~24.0 \pm ~6.0$ & 1.27 & 0.16 &
$2.32 \pm 0.5$ & 33.07 & 0.83 \cr
{}~~5 & 0.0 & $~7.4 \pm 0.1$ &$5.7 \pm 0.1$ & $~68.0 \pm 30.0$ & 1.32 & 0.17 &
$2.31 \pm 0.5$ & 33.20 & 0.81 \cr
&&&&&&&&& \cr
ISO & 0.5 & $~7.5 \pm 0.4$ &$5.7 \pm 0.2$ & $~24.0 \pm ~4.0$ & 1.33 & 0.13 &
$2.28 \pm 0.5$ & 32.89 & 0.85 \cr
{}~10 & 0.5 & $~7.6 \pm 0.4$ &$5.7 \pm 0.2$ & $~29.0 \pm ~8.0$ & 1.36 & 0.12 &
$2.28 \pm 0.5$ & 32.96 & 0.85 \cr
{}~~5 & 0.5 & $~7.8 \pm 0.1$ &$5.7 \pm 0.1$ & $~90.0 \pm 30.0$ & 1.34 & 0.16 &
$2.30 \pm 0.5$ & 33.15 & 0.82 \cr
&&&&&&&&& \cr
ISO & 1.0 & $~9.0 \pm 0.4$ &$5.6 \pm 0.2$ & $~33.0 \pm ~6.0$ & 1.47 & 0.09 &
$2.22 \pm 0.5$ & 32.77 & 0.87 \cr
{}~10 & 1.0 & $~9.3 \pm 0.2$ &$5.4 \pm 0.1$ & $~61.0 \pm 13.0$ & 1.39 & 0.17 &
$2.22 \pm 0.5$ & 32.85 & 0.86\cr
{}~~5 & 1.0 & $~8.4 \pm 0.1$ &$5.9 \pm 0.1$ & $~95.0 \pm 35.0$ & 1.48 & 0.09 &
$2.29 \pm 0.5$ & 33.07 & 0.83 \cr
&&&&&&&&& \cr
ISO & 1.5 & $11.0 \pm 0.2$ &$5.5 \pm 0.1$ & $~51.0 \pm ~4.0$ & 1.51 & 0.11 &
$2.18 \pm 0.5$ & 32.71 & 0.87 \cr
{}~10 & 1.5 & $10.6 \pm 0.2$ &$5.6 \pm 0.1$ & $~93.0 \pm 26.0$ & 1.30 & 0.20 &
$2.19 \pm 0.5$ & 32.81 & 0.86 \cr
{}~~5 & 1.5 & $~9.1 \pm 0.1$ &$6.3 \pm 0.1$ & $101.0 \pm 30.0$ & 1.94 & 0.01 &
$2.31 \pm 0.5$ & 33.01 & 0.84 \cr
&&&&&&&&& \cr
ISO & 2.0 & $13.4 \pm 0.2$ &$5.4 \pm 0.1$ & $~72.0 \pm ~5.0$ & 1.40 & 0.21 &
$2.16 \pm 0.5$ & 32.70 & 0.87 \cr
{}~10 & 2.0 & $11.8 \pm 0.3$ &$6.0 \pm 0.1$ & $105.0 \pm 36.0$ & 1.54 & 0.10 &
$2.23 \pm 0.5$ & 32.78 & 0.86 \cr
{}~~5 & 2.0 & $~9.9 \pm 0.1$ &$6.8 \pm 0.1$ & $102.0 \pm 15.0$ & 2.97 & 0.00 &
$2.35 \pm 0.5$ & 32.95 & 0.84 \cr
\spl}
\endinsert
\vfill\dosupereject

\pageinsert
\tabhl{\taone}{King-Michie - V Band Fitted Parameters}
\span\tcoli & \span\tcoll{.15}{9}{4} & \span\tcol{.15} & \span\tcol{.15} &
\span\tcol{.15} & \span\tcol{.15}
& \span\tcoll{.15}{9}{4} && \span\tcol{.15} \cr
\topperl

 & & & \multispan3 Photometry & & \multispan3 Velocities \cr
\tablerule
\hfil r$_a$ & x & W$_0$ & r$_s$ & c & $\chi_\nu^2$ & P($> \chi_\nu^2$) & v$_s$
& $\zeta^2$ & P($> \zeta^2$) \cr
\hfil (r$_s$) & &  & (pc) & & ($\nu=17$) & & (km s$^{-1}$) & & \cr
\tablerule

ISO &     & $~6.0 \pm 0.2$ & $4.3 \pm 0.2$ & $18.0 \pm ~2.0$ & 1.27 & 0.18 &
$2.35 \pm 0.5$ & 33.84 & 0.70 \cr
{}~10 &     & $~5.9 \pm 0.2$ & $4.4 \pm 0.2$ & $19.0 \pm ~3.0$ & 1.29 & 0.17 &
$2.35 \pm 0.5$ & 33.91 & 0.69 \cr
{}~~5 &     & $~5.7 \pm 0.2$ & $4.6 \pm 0.2$ & $21.0 \pm ~6.0$ & 1.37 & 0.12 &
$2.40 \pm 0.5$ & 34.10 & 0.65 \cr
&&&&&&&&& \cr
ISO & 0.0 & $~6.7 \pm 0.4$ & $5.5 \pm 0.2$ & $20.0 \pm ~4.0$ & 1.64 & 0.04 &
$2.35 \pm 0.5$ & 33.05 & 0.83 \cr
{}~10 & 0.0 & $~6.7 \pm 0.5$ & $5.5 \pm 0.2$ & $22.0 \pm ~7.0$ & 1.67 & 0.03 &
$2.35 \pm 0.5$ & 33.12 & 0.82 \cr
{}~~5 & 0.0 & $~7.3 \pm 0.4$ & $5.4 \pm 0.2$ & $65.0 \pm 30.0$ & 1.74 & 0.03 &
$2.35 \pm 0.5$ & 33.26 & 0.80 \cr
&&&&&&&&& \cr
ISO & 0.5 & $~7.3 \pm 0.5$ & $5.5 \pm 0.2$ & $22.0 \pm ~5.0$ & 1.71 & 0.03 &
$2.30 \pm 0.5$ & 32.93 & 0.85 \cr
{}~10 & 0.5 & $~7.2 \pm 0.6$ & $5.6 \pm 0.2$ & $24.0 \pm ~8.0$ & 1.73 & 0.03 &
$2.30 \pm 0.5$ & 33.01 & 0.84 \cr
{}~~5 & 0.5 & $~7.7 \pm 0.3$ & $5.5 \pm 0.2$ & $71.3 \pm 40.0$ & 1.72 & 0.03 &
$2.30 \pm 0.5$ & 33.19 & 0.81 \cr
&&&&&&&&& \cr
ISO & 1.0 & $~8.2 \pm 0.6$ & $5.7 \pm 0.3$ & $25.0 \pm ~6.0$ & 1.82 & 0.02 &
$2.30 \pm 0.5$ & 32.81 & 0.86 \cr
{}~10 & 1.0 & $~9.1 \pm 0.6$ & $5.3 \pm 0.3$ & $50.0 \pm 24.0$ & 1.79 & 0.02 &
$2.25 \pm 0.5$ & 32.87 & 0.86\cr
{}~~5 & 1.0 & $~8.4 \pm 0.2$ & $5.7 \pm 0.2$ & $94.5 \pm 35.0$ & 1.74 & 0.03 &
$2.30 \pm 0.5$ & 33.10 & 0.83 \cr
&&&&&&&&& \cr
ISO & 1.5 & $10.5 \pm 0.4$ & $5.5 \pm 0.2$ & $43.0 \pm ~6.0$ & 1.87 & 0.02 &
$2.20 \pm 0.5$ & 32.72 & 0.87 \cr
{}~10 & 1.5 & $10.4 \pm 0.4$ & $5.4 \pm 0.2$ & $73.0 \pm 35.0$ & 1.73 & 0.03 &
$2.20 \pm 0.5$ & 32.82 & 0.84 \cr
{}~~5 & 1.5 & $~9.1 \pm 0.1$ & $6.0 \pm 0.2$ & $89.3 \pm 35.0$ & 1.93 & 0.02 &
$2.30 \pm 0.5$ & 33.03 & 0.83 \cr
&&&&&&&&& \cr
ISO & 2.0 & $12.9 \pm 0.3$ & $5.3 \pm 0.2$ & $64.0 \pm ~8.0$ & 1.80 & 0.02 &
$2.20 \pm 0.5$ & 32.70 & 0.87 \cr
{}~10 & 2.0 & $11.7 \pm 0.2$ & $5.8 \pm 0.2$ & $93.0 \pm 25.0$ & 1.78 & 0.02 &
$2.25 \pm 0.5$ & 32.79 & 0.86 \cr
{}~~5 & 2.0 & $~9.9 \pm 0.1$ & $6.5 \pm 0.2$ & $99.0 \pm 17.0$ & 2.33 & 0.00 &
$2.35 \pm 0.5$ & 32.98 & 0.84 \cr
\spl
\endinsert
\vfill\dosupereject

\pageinsert
\baselineskip=13pt
\tabh{\taone}{King-Michie - B Band Derived Parameters}
\span\tcoli && \span\tcol{.05} \cr
\topper

\hfil         & &                   &               &                   &
        & \multispan2 Population & \multispan2 Dynamical \cr
\hfil r$_a$ & x & L$_{B0}$ & L$_B$ & $\rho_{0}$ & M & (M/L$_B$)$_0$ & (M/L$_B$)
& (M/L$_B$)$_0$ & (M/L$_B$) \cr
\hfil (r$_s$) & & (L$_B$\sol\ pc$^{-3}$) & (10$^5$L$_B$\sol) & (M\sol\
pc$^{-3}$) & (10$^5$M\sol) & (M/L$_B$)\sol & (M/L$_B$)\sol & (M/L$_B$)\sol &
(M/L$_B$)\sol \cr
\spacer

ISO &     & $200.0 \pm 10.0$ & $3.10 \pm 0.08$ & $52.0 \pm 20.0$ & $0.70 \pm
0.30$ &           &             & $0.21 \pm 0.08$ & $0.21 \pm 0.08$ \cr
{}~10 &     & $197.0 \pm 10.0$ & $3.14 \pm 0.08$ & $52.0 \pm 20.0$ & $0.70 \pm
0.30$ &           &             & $0.20 \pm 0.08$ & $0.20 \pm 0.08$ \cr
{}~~5 &     & $192.0 \pm 10.0$ & $3.21 \pm 0.11$ & $50.0 \pm 20.0$ & $0.70 \pm
0.30$ &           &             & $0.20 \pm 0.08$ & $0.20 \pm 0.08$ \cr
    &     &             &            &            &      &            &
   &            &            \cr
ISO & 0.0 & $182.0 \pm ~8.0$ & $3.30 \pm 0.12$ & $28.0 \pm 10.0$ & $1.20 \pm
0.45$ & 0.15  & 0.33 & $0.14 \pm 0.06$ & $0.34 \pm 0.15$ \cr
{}~10 & 0.0 & $178.0 \pm ~8.0$ & $3.32 \pm 0.13$ & $28.0 \pm 10.0$ & $1.20 \pm
0.45$ & 0.15  & 0.33 & $0.14 \pm 0.06$ & $0.34 \pm 0.15$ \cr
{}~~5 & 0.0 & $174.0 \pm 11.0$ & $3.64 \pm 0.07$ & $28.0 \pm 10.0$ & $1.45 \pm
0.50$ & 0.13  & 0.34 & $0.14 \pm 0.06$ & $0.38 \pm 0.15$ \cr
    &     &             &            &            &      &            &
   &            &            \cr
ISO & 0.5 & $178.0 \pm ~8.0$ & $3.33 \pm 0.13$ & $27.0 \pm 11.0$ & $1.40 \pm
0.50$ & 0.12  & 0.34 & $0.14 \pm 0.06$ & $0.40 \pm 0.15$ \cr
{}~10 & 0.5 & $177.0 \pm ~8.0$ & $3.37 \pm 0.14$ & $27.0 \pm 11.0$ & $1.45 \pm
0.50$ & 0.12  & 0.33 & $0.13 \pm 0.06$ & $0.40 \pm 0.15$ \cr
{}~~5 & 0.5 & $171.0 \pm 11.0$ & $3.58 \pm 0.07$ & $27.0 \pm 11.0$ & $1.75 \pm
0.50$ & 0.11 & 0.34 & 0$.14 \pm 0.06 $& 0$.44 \pm 0.20 $\cr
    &     &             &            &            &      &            &
   &            &            \cr
ISO & 1.0 & $178.0 \pm ~8.0$ & $3.50 \pm 0.14$ & $26.0 \pm 10.0$ & $2.10 \pm
0.85$ & 0.10 & 0.42 & $0.13 \pm 0.06$ & $0.55 \pm 0.25$ \cr
{}~10 & 1.0 & $185.0 \pm ~7.0$ & $3.61 \pm 0.09$ & $28.0 \pm 11.0$ & $2.35 \pm
0.95$ & 0.10 & 0.42 & $0.13 \pm 0.06$ & $0.61 \pm 0.25$ \cr
{}~~5 & 1.0 & $162.0 \pm 11.0$ & $3.48 \pm 0.08$ & $25.0 \pm 10.0$ & $2.10 \pm
0.85$ & 0.10 & 0.42 & $0.14 \pm 0.06$ & $0.57 \pm 0.25$ \cr
    &     &             &            &            &      &            &
   &            &            \cr
ISO & 1.5 & $183.0 \pm ~7.0$ & $3.64 \pm 0.09$ & $26.0 \pm 10.0$ & $3.55 \pm
1.40$ & 0.09 & 0.61 & $0.13 \pm 0.06$ & $0.91 \pm 0.40$ \cr
{}~10 & 1.5 & $173.0 \pm 10.0$ & $3.60 \pm 0.07$ & $26.0 \pm 10.0$ & $3.55 \pm
1.40$ & 0.09 & 0.60 & $0.13 \pm 0.06$ & $0.91 \pm 0.40$ \cr
{}~~5 & 1.5 & $150.0 \pm 10.0$ & $3.33 \pm 0.06$ & $22.0 \pm ~9.0$ & $2.80 \pm
1.10$ & 0.10 & 0.60 & $0.13 \pm 0.06$ & $0.78 \pm 0.30$ \cr
    &     &             &            &            &      &            &
   &            &            \cr
ISO & 2.0 & $183.0 \pm ~7.0$ & $3.70 \pm 0.08$ & $27.0 \pm 11.0$ & $6.15 \pm
2.45$ & 0.09 & 1.00 & $0.13 \pm 0.06$ & $1.54 \pm 0.65$ \cr
{}~10 & 2.0 & $157.0 \pm 10.0$ & $3.40 \pm 0.10$ & $23.0 \pm ~9.0$ & $4.95 \pm
1.95$ & 0.10 & 0.99 & $0.13 \pm 0.06$ & $1.35 \pm 0.55$ \cr
{}~~5 & 2.0 & $139.0 \pm ~8.0$ & $3.17 \pm 0.06$ & $20.0 \pm ~8.0$ & $3.70 \pm
1.45$ & 0.14 & 0.98 & $0.11 \pm 0.05$ & $1.07 \pm 0.45$ \cr
\sp
\endinsert
\vfill\dosupereject

\pageinsert
\baselineskip=13pt
\tabh{\taone}{King-Michie - V Band Derived Parameters}
\span\tcoli && \span\tcol{.05} \cr
\topper

\hfil         & &                   &               &                   &
        & \multispan2 Population & \multispan2 Dynamical \cr
\hfil r$_a$ & x & L$_{V0}$ & L$_V$ & $\rho_{0}$ & M & (M/L$_V$)$_0$ & (M/L$_V$)
& (M/L$_V$)$_0$ & (M/L$_V$) \cr
\hfil (r$_s$) & & (L$_V$\sol\ pc$^{-3}$) & (10$^5$L$_V$\sol) & (M\sol\
pc$^{-3}$) & (10$^5$M\sol) &  (M/L$_V$)\sol & (M/L$_V$)\sol & (M/L$_V$)\sol &
(M/L$_V$)\sol \cr
\spacer

ISO &     & $222.0 \pm 14.0$ & $3.07 \pm 0.09$ & $48.0 \pm 20.0$ & $0.65 \pm
0.25$ &           &             & $0.20 \pm 0.08$ & $0.20 \pm 0.08$ \cr
{}~10 &     & $222.0 \pm 14.0$ & $3.08 \pm 0.09$ & $48.0 \pm 20.0$ & $0.65 \pm
0.25$ &           &             & $0.20 \pm 0.08$ & $0.20 \pm 0.08$ \cr
{}~~5 &     & $213.0 \pm 14.0$ & $3.10 \pm 0.14$ & $46.0 \pm 20.0$ & $0.65 \pm
0.25$ &           &             & $0.20 \pm 0.08$ & $0.20 \pm 0.08$ \cr
    &     &             &            &            &      &            &
   &            &            \cr
ISO & 0.0 & $200.0 \pm 11.0$ & $3.13 \pm 0.14$ & $30.0 \pm 12.0$ & $1.10 \pm
0.45$ & 0.18  & 0.40 & $0.12 \pm 0.06$ & $0.32 \pm 0.14$ \cr
{}~10 & 0.0 & $198.0 \pm 11.0$ & $3.14 \pm 0.14$ & $30.0 \pm 12.0$ & $1.10 \pm
0.45$ & 0.18  & 0.40 &$0.12 \pm 0.06$ &$0.30 \pm 0.14$ \cr
{}~~5 & 0.0 & $202.0 \pm 12.0$ & $3.41 \pm 0.14$ & $30.0 \pm 12.0$ & $1.30 \pm
0.50$ & 0.16  & 0.40 &$0.12 \pm 0.06$ &$0.35 \pm 0.14$ \cr
    &     &             &            &            &      &            &
   &            &            \cr
ISO & 0.5 & $198.0 \pm 11.0$ & $3.16 \pm 0.14$ & $30.0 \pm 12.0$ & $1.30 \pm
0.50$ & 0.15  & 0.40 &$0.13 \pm 0.06$ &$0.37 \pm 0.14$ \cr
{}~10 & 0.5 & $195.0 \pm 11.0$ & $3.16 \pm 0.16$ & $30.0 \pm 12.0$ & $1.30 \pm
0.50$ & 0.14  & 0.41 &$0.13 \pm 0.06$ &$0.37 \pm 0.14$ \cr
{}~~5 & 0.5 & $197.0 \pm 13.0$ & $3.38 \pm 0.12$ & $30.0 \pm 12.0$ & $1.60 \pm
0.50$ & 0.13 & 0.41 &$0.13 \pm 0.06$ &$0.42 \pm 0.19$ \cr
    &     &             &            &            &      &            &
   &            &            \cr
ISO & 1.0 & $192.0 \pm 11.0$ & $3.19 \pm 0.14$ & $26.0 \pm 10.0$ & $1.70 \pm
0.70$ & 0.13 & 0.50 &$0.12 \pm 0.05$ &$0.48 \pm 0.19$ \cr
{}~10 & 1.0 & $203.0 \pm 11.0$ & $3.39 \pm 0.17$ & $30.0 \pm 12.0$ & $2.15 \pm
0.85$ & 0.12 & 0.51 &$0.13 \pm 0.06$ &$0.56 \pm 0.24$ \cr
{}~~5 & 1.0 & $183.0 \pm 14.0$ & $3.34 \pm 0.11$ & $27.0 \pm 11.0$ & $2.05 \pm
0.85$ & 0.12 & 0.51 &$0.13 \pm 0.06$ &$0.55 \pm 0.24$ \cr
    &     &             &            &            &      &            &
   &            &            \cr
ISO & 1.5 & $198.0 \pm 10.0$ & $3.37 \pm 0.14$ & $27.0 \pm 11.0$ & $3.05 \pm
1.20$ & 0.11 & 0.75 &$0.12 \pm 0.05$ &$0.80 \pm 0.31$ \cr
{}~10 & 1.5 & $198.0 \pm 11.0$ & $3.39 \pm 0.14$ & $27.0 \pm 11.0$ & $3.20 \pm
1.25$ & 0.11 & 0.74 &$0.12 \pm 0.05$ &$0.83 \pm 0.31$ \cr
{}~~5 & 1.5 & $170.0 \pm 13.0$ & $3.22 \pm 0.09$ & $25.0 \pm 10.0$ & $2.70 \pm
1.10$ & 0.12 & 0.74 &$0.13 \pm 0.06$ &$0.75 \pm 0.28$ \cr
    &     &             &            &            &      &            &
   &            &            \cr
ISO & 2.0 & $202.0 \pm ~9.0$ & $3.48 \pm 0.14$ & $28.0 \pm 11.0$ & $5.40 \pm
2.15$ & 0.11 & 1.19 &$0.12 \pm 0.05$ &$1.38 \pm 0.57$ \cr
{}~10 & 2.0 & $181.0 \pm 11.0$ & $3.29 \pm 0.10$ & $25.0 \pm 10.0$ & $4.55 \pm
1.80$ & 0.12 & 1.20 &$0.12 \pm 0.05$ &$1.24 \pm 0.47$ \cr
{}~~5 & 2.0 & $160.0 \pm 10.0$ & $3.08 \pm 0.09$ & $22.0 \pm ~9.0$ & $3.50 \pm
1.40$ & 0.14 & 1.22 &$0.12 \pm 0.05$ &$1.02 \pm 0.43$ \cr
\sp
\endinsert
\vfill\dosupereject

\pageinsert
\tabh{\taone}{Radial Velocities}
\span\tcoli && \span\tcol{.3} \cr
\topper

\hfil ID & Telescope & \Rc2 & $\Theta$ & v$_r$ & $\left<\hbox{v}_r\right>$ &
HJD & V & B-V \cr
\hfil  & & (pc) & ($^\circ$) & (km s$^{-1}$) & (km s$^{-1}$) & (--2448000) &
(mag) & (mag) \cr
\spacer

  1 & CTIO &  1.4 & 346.7 &$295.0 \pm 2.1$ &$294.9 \pm 1.8$ & 276.7601 & 16.88
& 0.82 \cr
    & CTIO &      &       &$294.6 \pm 3.3$ &                 & 275.7578 & & \cr
  2 & CTIO &  1.5 & 273.3 &$296.4 \pm 2.1$ &$294.8 \pm 1.0$ & 276.7251 & 15.94
& 1.36 \cr
    & LCO~ &      &       &$294.6 \pm 1.2$ &                 & 287.5633 & & \cr
    & CTIO &      &       &$293.3 \pm 2.2$ &                 & 275.7289 & & \cr
  3 & CTIO &  2.2 & 119.1 &$292.4 \pm 2.0$ &$292.8 \pm 1.5$ & 277.5507 & 15.89
& 1.38 \cr
    & CTIO &      &       &$293.4 \pm 2.5$ &                 & 275.8156 & & \cr
  4 & LCO~ &  2.2 & 316.6 &$294.8 \pm 1.5$ &$294.2 \pm 1.2$ & 288.5598 & 16.31
& 1.26 \cr
    & CTIO &      &       &$293.0 \pm 2.5$ &                 & 276.7102 & & \cr
    & CTIO &      &       &$293.5 \pm 2.9$ &                 & 275.7175 & & \cr
  5 & CTIO &  2.4 & 353.7 &$291.0 \pm 2.9$ &$292.8 \pm 1.6$ & 275.7691 & 16.09
& 1.01 \cr
    & CTIO &      &       &$293.6 \pm 1.9$ &                 & 276.7798 & & \cr
  6 & LCO~ &  2.5 & ~89.0 &$291.4 \pm 1.3$ &$291.4 \pm 1.3$ & 287.6484 & 16.35
& 1.08 \cr
  7 & CTIO &  3.0 & 139.4 &$291.4 \pm 1.7$ &$291.2 \pm 1.4$ & 277.5654 & 16.48
& 1.44 \cr
    & CTIO &      &       &$290.8 \pm 2.4$ &                 & 275.8273 & & \cr
  8 & CTIO &  3.4 & ~44.4 &$293.4 \pm 3.3$ &$296.3 \pm 2.2$ & 275.7979 & 16.92
& 1.03 \cr
    & CTIO &      &       &$298.6 \pm 2.9$ &                 & 277.7268 & & \cr
  9 & CTIO &  3.5 & 284.0 &$291.4 \pm 2.2$ &$291.4 \pm 2.2$ & 277.6689 & 16.35
& 1.54 \cr
 10 & CTIO &  3.6 & 358.1 &$293.6 \pm 1.9$ &$293.6 \pm 1.9$ & 277.7086 & 16.57
& 1.01 \cr
 11 & LCO~ &  3.9 & 327.2 &$296.7 \pm 1.0$ &$296.7 \pm 1.0$ & 287.5977 & 17.09
& 1.39 \cr
 12 & LCO~ &  4.0 & ~~4.2 &$292.9 \pm 1.4$ &$292.9 \pm 1.4$ & 287.6338 & 16.57
& 1.01 \cr
 13 & CTIO &  4.1 & ~36.7 &$304.3 \pm 2.6$ & Not Used \hfill & 276.7952 & 16.11
& 1.52 \cr
    & CTIO &      &       &$292.7 \pm 3.6$ &                 & 275.7854 & & \cr
 14 & CTIO &  4.3 & ~16.1 &$290.7 \pm 2.1$ &$290.7 \pm 2.1$ & 277.6874 & 16.59
& 1.46 \cr
 15 & LCO~ &  4.3 & 234.7 &$295.2 \pm 1.5$ &$295.2 \pm 1.5$ & 287.5769 & 16.75
& 1.25 \cr
\sp
\endinsert
\vfill\dosupereject

\pageinsert
\tabh{\taonen}{Radial Velocities}
\span\tcoli && \span\tcol{.3} \cr
\topper

\hfil ID & Telescope & \Rc2 & $\Theta$ & v$_r$ & $\left<\hbox{v}_r\right>$ &
HJD & V & B-V \cr
\hfil  & & (pc) & ($^\circ$) & (km s$^{-1}$) & (km s$^{-1}$) & (--2448000) &
(mag) & (mag) \cr
\spacer

 16 & CTIO &  4.3 & 340.1 &$296.2 \pm 2.8$ &$291.9 \pm 0.9$ & 275.7405 & 16.02
& 1.38 \cr
    & LCO~ &      &       &$291.3 \pm 1.1$ &                 & 302.5609 & & \cr
    & CTIO &      &       &$291.4 \pm 2.0$ &                 & 276.7448 & & \cr
 17 & LCO~ &  4.6 & 120.1 &$295.5 \pm 2.1$ &$295.5 \pm 2.1$ & 289.6231 & 16.98
& 1.10 \cr
 18 & CTIO &  5.0 & 308.3 &$294.2 \pm 3.4$ &$294.2 \pm 3.4$ & 277.6038 & 16.70
& 1.42 \cr
 19 & LCO~ &  5.0 & 104.0 &$292.6 \pm 1.5$ &$292.6 \pm 1.5$ & 289.6057 & 16.90
& 1.04 \cr
 20 & LCO~ &  5.2 & 172.1 &$292.8 \pm 0.9$ &$292.8 \pm 0.9$ & 289.5814 & 16.96
& 1.27 \cr
 21 & CTIO &  5.3 & 346.8 &$297.5 \pm 2.6$ &$297.5 \pm 2.6$ & 277.7824 & 16.76
& 1.25 \cr
 22 & CTIO &  5.4 & 163.2 &$294.4 \pm 1.8$ &$294.4 \pm 1.8$ & 277.8052 & 16.70
& 1.34 \cr
 23 & LCO~ &  5.6 & ~21.7 &$291.8 \pm 1.6$ &$291.8 \pm 1.6$ & 287.6130 & 16.64
& 1.31 \cr
 24 & CTIO &  5.6 & ~70.3 &$291.8 \pm 1.7$ &$291.8 \pm 1.7$ & 277.5862 & 16.31
& 1.11 \cr
 25 & LCO~ &  8.1 & ~67.5 &$291.1 \pm 1.6$ &$291.1 \pm 1.6$ & 288.6356 & 16.87
& 1.51 \cr
 26 & CTIO &  8.4 & 325.4 &$291.3 \pm 2.9$ &$291.3 \pm 2.9$ & 277.6291 & 17.02
& 1.52 \cr
 27 & LCO~ &  8.9 & 142.4 &$290.6 \pm 0.9$ &$290.6 \pm 0.9$ & 288.6536 & 16.35
& 1.60 \cr
 28 & LCO~ &  9.1 & ~~2.8 &$290.1 \pm 1.3$ &$290.1 \pm 1.3$ & 288.5932 & 16.91
& 1.41 \cr
 29 & LCO~ &  9.5 & ~40.8 &$296.6 \pm 1.3$ &$296.6 \pm 1.3$ & 288.6113 & 16.90
& 1.64 \cr
 30 & LCO~ & 11.1 & ~11.3 &$297.3 \pm 0.9$ &$297.4 \pm 0.7$ & 288.5716 & 15.79
& 1.89 \cr
    & LCO~ &      &       &$297.6 \pm 1.0$ &                 & 287.6762 & & \cr
    & CTIO &      &       &$296.7 \pm 3.2$ &                 & 275.8437 & & \cr
 31 & LCO~ & 12.1 & 133.7 &$295.3 \pm 1.0$ &$295.3 \pm 1.0$ & 289.5633 & 16.56
& 1.77 \cr
 32 & LCO~ & 12.5 & ~86.6 &$295.7 \pm 1.2$ &$295.7 \pm 1.2$ & 289.6661 & 16.82
& 1.56 \cr
 33 & LCO~ & 12.7 & 285.1 &$291.9 \pm 1.4$ &$291.9 \pm 1.4$ & 289.6897 & 16.68
& 1.74 \cr
 34 & LCO~ & 12.7 & 116.2 &$290.3 \pm 1.3$ &$290.3 \pm 1.3$ & 289.6481 & 16.89
& 1.22 \cr
 35 & LCO~ & 14.0 & 325.4 &$287.5 \pm 1.9$ &$287.5 \pm 1.9$ & 289.7078 & 16.84
& 2.10 \cr
 36 & LCO~ & 19.9 & 155.8 &$294.0 \pm 1.4$ &$294.0 \pm 1.4$ & 302.6129 & 16.66
& 1.80 \cr
\sp
\endinsert
\vfill\dosupereject
\pageinsert
\baselineskip=13pt
\tabhl{\taone}{Oblate Spheroid Models}
\span\tcolil{9}{4} & \span\tcol{.3} & \span\tcol{.3} & \span\tcol{.3} &
\span\tcoll{.3}{9}{4} && \span\tcol{.3} \cr
\topperl

\hfil & \multispan8 \hfil Non-Rotating \hfil \vvbs{9}{4} \hfil \hskip3.2pt
\phantom{No}Rotating\phantom{n-} \hskip3.3pt\hfil \cr
\tablerule
\hfil $\hbox{R}_a$ & M$_\infty$ & $\zeta^2$ & P($> \zeta^2$) & M/L$_V$ &
M$_\infty$ & $\zeta^2$ & P($> \zeta^2$) & M/L$_V$ \cr
\hfil (pc)  & ($10^4$M$_\odot$) & & & (M$_\odot$/L$_\odot$) & ($10^4$M$_\odot$)
& & &  (M$_\odot$/L$_\odot$) \cr
\tablerule

{}~$\infty$&$6.5 \pm 2.5$ & 33.75 & 0.72 &$0.20 \pm 0.08$ &$7.6 \pm 2.5$ &
39.14 & 0.23 &$0.23 \pm 0.09$ \cr
   100 &$6.5 \pm 2.5$ & 33.76 & 0.71 &$0.20 \pm 0.08$ &$7.6 \pm 2.5$ & 39.28 &
0.23 &$0.23 \pm 0.09$ \cr
   ~50 &$6.5 \pm 2.5$ & 33.82 & 0.71 &$0.20 \pm 0.08$ &$7.8 \pm 2.5$ & 39.69 &
0.22 &$0.24 \pm 0.09$ \cr
   ~25 &$6.5 \pm 2.5$ & 34.07 & 0.66 &$0.20 \pm 0.08$ &$7.9 \pm 3.0$ & 41.14 &
0.18 &$0.24 \pm 0.09$ \cr
   ~10 &$6.6 \pm 2.5$ & 35.41 & 0.44 &$0.20 \pm 0.08$ &$9.0 \pm 3.0$ & 48.04 &
0.09 &$0.27 \pm 0.10$ \cr
   ~~5 &$7.1 \pm 2.5$ & 36.98 & 0.30 &$0.22 \pm 0.09$ &$9.8 \pm 3.5$ & 59.18 &
0.03 &$0.29 \pm 0.10$ \cr
\spl
\endinsert

\bye